\documentstyle[amssym,emulateapj]{article} 

\def\ls{{_<\atop^{\sim}}}
\def\gs{{_>\atop^{\sim}}}
\def\cgs{{ erg cm$^{-2}$ s$^{-1}$}}
\begin{document}

\title{Probing the warm Inter-Galactic Medium through 
absorption against Gamma Ray Bursts X-ray afterglows}

\author{F. Fiore\altaffilmark{1}, F. Nicastro\altaffilmark{1,2}, 
S. Savaglio\altaffilmark{1},
L. Stella\altaffilmark{1} \& M. Vietri\altaffilmark{3}}

\altaffiltext{1} {Osservatorio Astronomico di Roma, Via Frascati 33,
I-00044 Monteporzio Catone, Italy}
\altaffiltext{2} {Harvard-Smithsonian Center of Astrophysics, 
60 Garden Street,  Cambridge MA 02138 USA}
\altaffiltext{3} {Dipartimento di Fisica, Universit\`a Roma Tre, 
via della Vasca Navale 84, I-00146, Roma, Italy}

%
\submitted{version: 5-September-2000}

\begin{abstract}
Gamma Ray Burst (GRB) afterglows close to their peak intensity are
among the brightest X-ray sources in the sky.  Despite their fast
power-law like decay, when fluxes are integrated from minutes up to
hours after the GRB event, the corresponding number counts (logN-logF
relation) far exceeds that of any other high redshift (z$>0.5$)
source, the flux of which is integrated over the same time
interval. We discuss how to use X--ray afterglows of GRBs as distant
beacons to probe the warm ($10^5 K<T<10^7 K$) intergalactic matter in
filaments and outskirts of clusters of galaxies by
means of absorption features, the ``X-ray forest''. According to
current cosmological scenarios this matter may comprise 
$30-40\%$ of the baryons in the Universe at z$<1$.
Present-generation X-ray spectrometers such as those on Chandra and
XMM-Newton can detect it along most GRBs' lines of sight, provided
afterglows are observed fast enough (within hours) after the burst.  A
dedicated medium-sized X-ray telescope (effective area $\ls0.1$ m$^2$)
with pointing capabilities similar to that of Swift (minutes) and high
spectral resolution ($E/\Delta E \gs 300$) would be very well suited
to exploit the new diagnostic and study the physical conditions in the
Universe at the critical moment when structure is being formed.
\end{abstract}


\section{Introduction}

Gamma Ray Burst (GRB) afterglows carry a relatively large fraction of
the total GRB flux, often exceeding the total energy budget of the
main event itself.  Minutes after the GRB events, they are by far the
brighest sources in the sky at cosmological redshifts (the redshift of
9 GRBs has been measured so far, ranging between z=0.4 and z=3.4, with
a median of about z=1.3).  For this reason the optical and infrared
GRB afterglows have been proposed as probes of the high redshift
Universe through the detection of absorption line systems along the
line of sight (Lamb \& Reichart 2000, Ciardi \& Loeb 2000). 
However, optical and infrared afterglows have been detected
in less than half of the GRBs observed by BeppoSAX. Moreover, they
only carry a small fraction of the total GRB afterglow flux, most of
which is in the X-ray band. We propose here to exploit X-ray resonant
scattering lines in the GRB afterglow X-ray spectra to probe the warm
component of the inter-galactic medium (IGM).

Hydrodynamic simulations show that at z$<1$ a large fraction
($30-40\%$) of the baryons in the Universe are in a warm phase,
shock-heated to temperatures of $10^5-10^7$ K during the collapse of
density perturbations (warm phase hereafter, see Dav\`e et al. 2000
and references therein).  According to the same simulations $10-20\%$ of
the remaining baryons end up in clusters of galaxies (with $T>10^7$ K,
hot phase), and $30-40\%$ are in stars and colder gas clouds ($T<10^5$
K, cold phase).  Both the hot and the cold phases of the IGM have been
detected and studied in detail in the X-ray and O-UV bands
respectively.  The cold gas is revealed through UV rest frame
absorption lines.  At z$>1.5$ these lines are redshifted in the
optical and so easily detected in the spectra of bright background
sources like quasars. At z$\ls1.5$ their study is complicated by the
limited capabilities of UV instruments.  Hot gas shines in the 0.1--10
keV band due to bremsstrahlung emission and line emission, both
proportional to the square of the gas density, and so primarily
tracing peak densities. On the other hand, observations of the warm
IGM expected away from the high density regions have yielded so far
only limited information.  Gas with $T\approx10^5$ K has been revealed
through OVI absorption at 1032, 1038 \AA~ at z=0.1--0.3 (Tripp et al.
2000, Tripp \& Savage, 2000). OVI has been detected at
z=2-4 too (Kirkman \& Tytler 1997, 1999), but in these cases the OVI
lines are probably associated with Lyman-limit systems
($16<$logN(HI)$<19$) and therefore with colder gas clouds in galaxies.
Warmer IGM ($10^6<T<10^7$ K) can be detected and studied by measuring
both the photoelectric absorption edges and the resonant scattering
lines of C, O, Ne, Si, S, Mg, Fe, etc. in the X-ray spectra of bright
background objects (the X-ray Gunn-Peterson test, Sherman \& Silk
1979, Shapiro \& Bahcall 1980, Aldcroft et al. 1994, Markevitch 1999).
X-ray spectroscopy of absorption edges and resonant scattering lines
is the {\it only} tool to probe such a warm, low density gas.  In fact
the strength of these features is linearly proportional to the gas
density and therefore below a critical value, absorption wins over
emission. This can be done best by studying with adequate energy
resolution (E/$\Delta E\gs 300$) a bright background source. Previous
studies proposed quasars as background objects (Aldcroft et al. 1994,
Hellsten et al. 1998, Perna \& Loeb 1998).  The problem is that X-ray
bright z$\gs0.3$ quasars are rare: only 9 of the 96 AGNs in the HEAO1
Grossan et al. (1992) sample (flux of
$S_{2-10keV}\gs1.5\times10^{-11}$ \cgs) have z$>0.3$ and only 3 have
z$>1$. Only these AGNs can provide a few~$\times10^4$ counts in deep
($\approx100$ ks) exposures with present-generation X-ray satellites
(Chandra and XMM-Newton).  The systematic study of the warm IGM using
fainter AGNs must then await for missions of the size of
Constellation-X\footnote{see e.g.
http://constellation.gsfc.nasa.gov/science/igm2.html} and
XEUS\footnote{ftp://astro.estec.esa.nl/pub/XEUS/BROCHURE/brochure.ps.gz and
http://www.ias.rm.cnr.it/sax/probing.html}
(collecting area of $1-10$~m$^2$), foreseen for the next decade.  We
find instead that {\it every} GRB X-ray afterglow can provide enough
counts to study the warm IGM by using medium-high resolution X-ray
spectroscopy, if observed within a few minutes after the GRB event.
In just one year of activity an instrument with the effective area of
a XMM-Newton or Astro-E mirror unit and a high resolution focal plane
detector (such as a grating or a calorimeter), may study accurately
tens to hundreds of lines of sight.  The study of the warm filaments
and cluster outskirts, possibly complemented with UV absorption
studies, will provide new crucial information on the phase at which
galaxies, groups and clusters formed, the metal enrichment and heating
histories of the IGM, and the feedback between hot and warm halos and
star-formation in galaxies (see e.g. Cavaliere et al. 2000).

\section{The X-ray logN-logF of GRBs and other high z sources }

The study of afterglows starting immediately after the GRB event is
the main goal of missions such as Swift\footnote{Swift will be
launched in 2003, see http://swift.sonoma.edu/}. Briefly, once a GRB
is detected, the satellite is automatically maneuvred to begin imaging
within tens of seconds the GRB region with optical and X-ray
telescopes. We derive in this section the expected high Galactic
latitude number counts (logN-log$F_{40}(t)$) of the X-ray afterglows,
as a function of the flux integrated starting from 40 seconds after
the GRB peak up to a time $t$ (the afterglow fluence $F_{40}(t)$). We
focus on the soft X-ray band since the main absorption features
imprinted by the IGM on the GRB X-ray spectra are below $\sim1$ keV
(see below).

Frontera et al. (2000a,b) studied the evolution of the
$\gamma-$ray and X--ray spectra of 8 GRBs seen by BeppoSAX.  Using
their spectra and light curves we computed the ratio between the 2-10
keV flux 30-40 seconds after the GRB peak and the 50-300
keV flux at the peak. All points but one are in the range
0.02-0.2. The outlier, GRB980329, has a ratio ten time smaller, and is
the only GRB in the Frontera et al. (2000b) sample to show significant
absorption in the 2-20 keV spectrum (see below).  In the following
estimates we conservatively assume a ratio of 0.01.

GRB X-ray afterglows decay with time as power laws with index $\delta$
in the range $-1\div -1.5$ (Costa 1999). Substantial evidence supports
the view that the X-ray afterglow starts well within the GRB duration,
when the GRB spectrum softens markedly. Indeed, when the power law
decay of the X-ray afterglow observed hours to days after several GRBs
is extrapolated backward in time, a good matching is found with the
BeppoSAX WFC flux a few tens of seconds after burst onset (Piro 1999,
Frontera et al. 2000b) We are thus justified in estimating the fluence
of the X-ray afterglow by integrating the X-ray emission starting 40
sec after the GRB peak. We adopt a total integration time of $40\;$ks,
assuming that up to this time after the main event the high--frequency
cut--off of the synchrotron spectrum has not yet moved into the X--ray
band. This is consistent with the fact that $\approx90\%$ of the
bursts do display an X--ray afterglow (the few non-detection may be
due to an insufficient sensitivity and/or too long a delay in the
BeppoSAX NFI follow--ups). The X-ray spectrum deduced from both the
WFC data, acquired seconds to minutes after the burst, and the LECS
and MECS spectra, acquired several hours after the burst, are
consistent with a power law model with energy index
$\alpha_E\approx1.0$. So far, there is evidence for intrinsic X-ray
absorption in one GRB only: GRB980329, $N_H\approx10^{22}$ cm$^{-2}$,
assuming solar abundances (Frontera et al. 2000b, Owens et al. 1998).
All other GRBs analyzed by Owens et al. (1998) have a measured column
density at the GRB redshift, where available, consistent with 0 at the
2 $\sigma$ level.  Excluding the GRB980329 event, the average best fit
$N_H$ at the GRB redshifts is $2.5\times10^{21}$ cm$^{-2}$.
Since the column density computed in the observer frame scales roughly
as $(1+z)^{-8/3}$ the spectrum seen by a z=0.1-0.3 IGM cloud has an
effective cutoff corresponding to an $N_H\approx0.5$ times that
obtained at the mean GRB redshift of 1.  We conservatively assume in
the following a GRB $N_H$ in the range $1-3\times10^{21}$
cm$^{-2}$ {\it as seen by a gas cloud at z$\ls0.3$}.

\begin{figure*}[t]
\plottwo{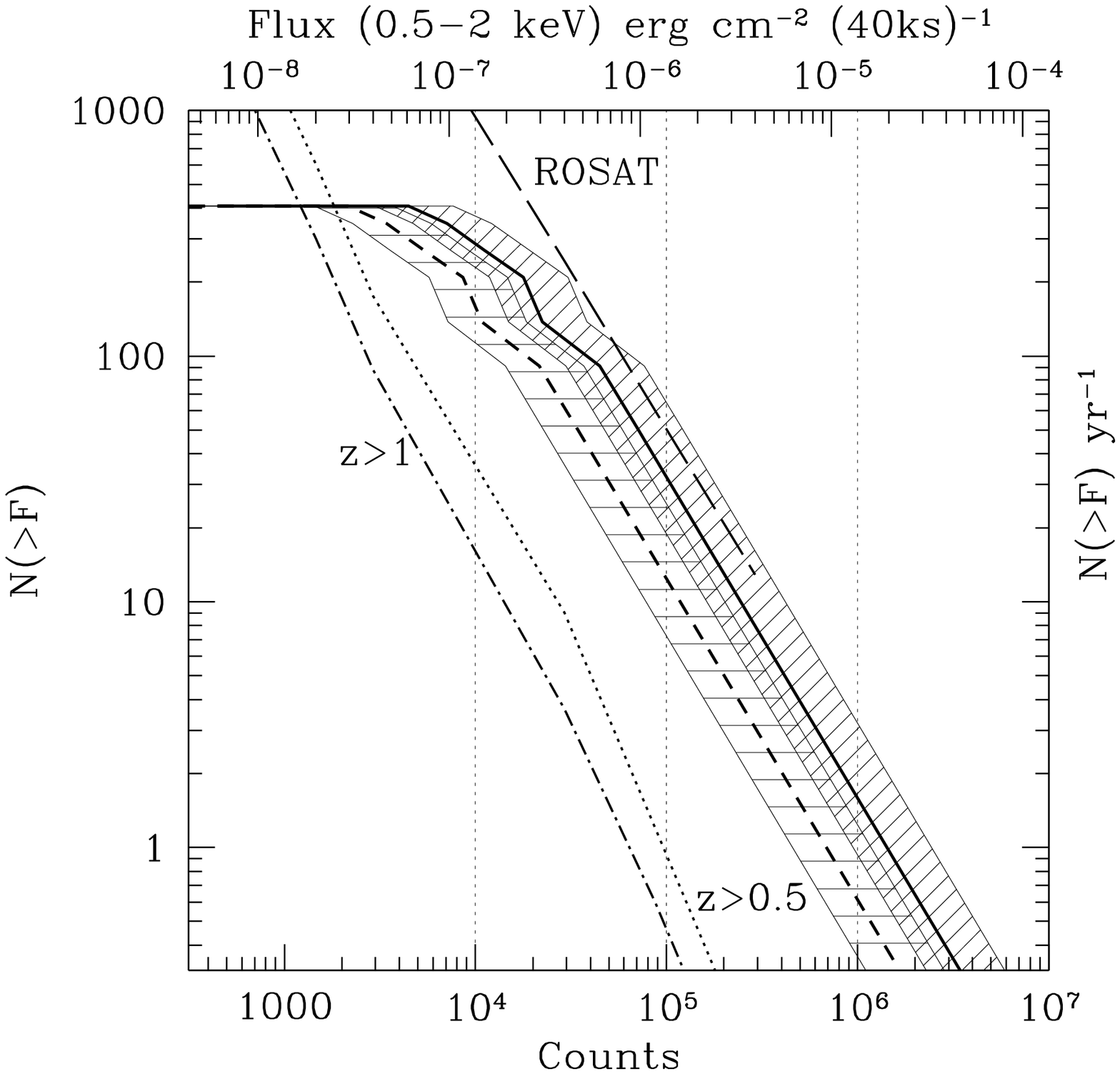}{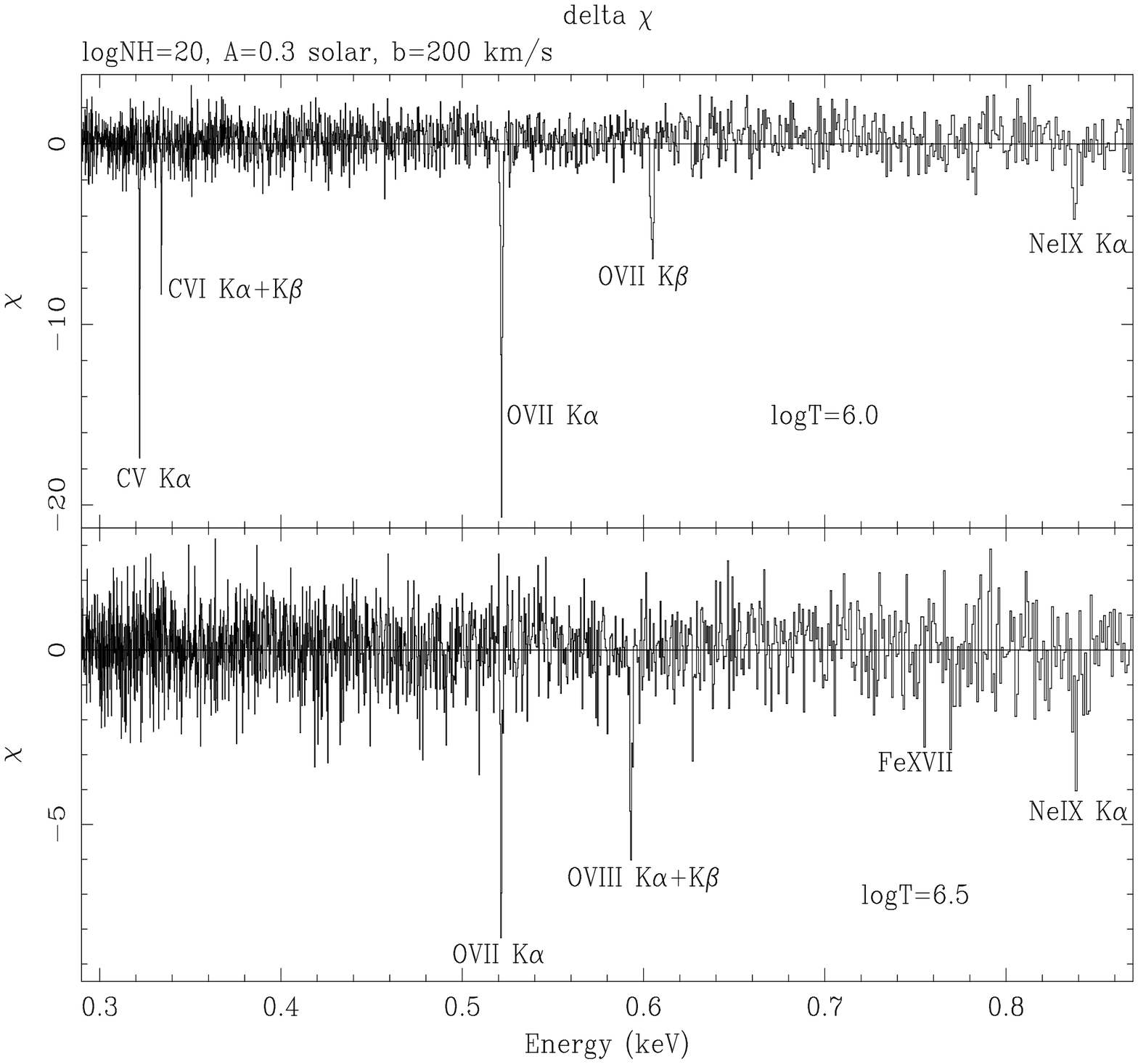}
\caption{ 
The number of GRBs at high Galactic latitude versus the 0.5-2 keV
fluence, i.e. the flux integrated in 40 ks (upper x-axis), and the
corresponding counts accumulated in 40 ks by an instrument with the
throughput of the XMM-RGS (one unit, lower x-axis). Fast GRBs
(duration of $\ls3$ s) contribute to $\sim30\%$ of the total.  An
intrinsic GRB $N_H$ of $10^{21}$ cm$^{-2}$ (thick solid line) and
$3\times10^{21}$ cm$^{-2}$ (thick dashed line) {\it as seen at redshift
$\ls0.3$} is assumed.  These lines are computed for a power law
GRB decay index of $\delta=-1.3$. The shaded bands surrounding each 
thick line have  $\delta=-1.1$ as the right hand boundary and
$\delta=-1.5$ as the left hand boundary. The
long dashed line is the ROSAT logN-logF.  The dotted line and the
dot-dashed line show the logN-logF of the z$>0.5$ and z$>1$ ROSAT
sources respectively.  Vertical dotted lines mark spectra with
$10^{4}$, $10^{5}$ and $10^{6}$ counts respectively.
Fig. 2.-- Residuals after subtraction of the best fit continuum from
simulations of grating spectra 
trasmitted by a cloud of gas at z=0.1 and of 
$N_H=10^{20}$ cm$^{-2}$, metal abundances $0.3 Z_\odot$, 
$b=200$ km/s and $T=10^6$ and $10^{6.5}$ K. See the text for details}
\label{counts}
\end{figure*}

The GRB peak flux $\log N- \log S$ relationship is well known from
$10^{-8}$ to $10^{-4}$ \cgs (50-300 keV, see e.g. Fishman \& Meegan
1995).  Using the relationship between the GRB peak flux and the 2-10
keV emission 30-40 seconds after the peak and assuming the X-ray
spectral shape discussed above, we predict the logN-logS of the 0.5-2
keV X-ray emission as it will be first seen by satellites with
reaction capabilities similar to those of Swift. Figure 1 shows the
number of objects at high galactic latitude (i.e. in half of the sky)
as a function of the 0.5-2 keV flux integrated in 40 ks (the fluence
F, upper scale) and of the total counts accumulated in 40 ks by an
instrument with the effective area of the XMM-Newton RGS (one
unit). The GRB logN-logF is compared to the ROSAT logN-logF (adapted
from Hasinger et al. 1998).  This is dominated at high fluxes by {\it
nearby} Seyfert galaxies and BL Lacertae objects.  Most of the IGM is
beyond the brightest 0.5-2 keV sources in the sky. The dotted line and
the dot-dashed line show the number of 0.5-2 keV sources with z$>0.5$
and z$>1$ respectively, as estimated using the soft X-ray logN-logS of
bright blazars (e.g. Wolter et al. 1991), the fraction of identified
AGN in the RASS and the results of synthetic AGN models for the X-ray
background (Comastri et al 1995).  Figure 1 shows that under our
conservative assumption there should be a few lines of sight per month
with $\gs10^5$ counts.  A few lines of sight per year may be studied
with more than $10^6$ counts.  These estimates have to be compared
with $\approx10$ lines of sight in total which can be studied with
$\gs10^5$ counts by using z$>0.5$ AGNs as background beacons.  We
remark that this result strongly depends on the speed with which the
afterglow is observed, i.e. for $\delta=-1.3$ twice as many photons
are emitted between $t_0$ and $10t_0$ than are emitted between $10t_0$
and $100t_0$.

\section{X-ray absorption features from the IGM}

In order to study the warm IGM with background beacons, metal
abundances must be sufficiently high that relatively strong absorption
features are produced.  Using numerical simulations Cen \& Ostriker
(1999) estimate a metallicity of at least $0.1 Z_\odot$ for z$<0.5$,
depending of the gas density.  This is supported by several pieces of
evidence.  First, the metallicity of the hot intracluster medium is
found to be $\approx0.3 Z_\odot$ in many cases; Second, collisionally
excited OVI lines have been detected at z$<0.3$ (Tripp et al. 2000,
Tripp \& Savage 2000).  Third, CIV and SiIV associated with Ly$\alpha$
clouds have been detected at z$>1.5$ (e.g. Songaila \& Cowie 1996,
Ellison et al., 1999) Finally, the nine Damped Ly$\alpha$ systems with
detected metals and z$<0.7$ have a metallicity in the range $0.4-0.6
Z_\odot$ (Savaglio et al. 2000), indicating that
galaxies at z$<0.7$ can provide metals to the IGM.

The IGM transmitted spectrum can be used as a diagnostic of
temperature, metal abundances, column and volume densities and gas
dynamics. We calculated the line strength and profiles expected from a
cloud of gas at a given redshift, with given temperature, column
density and metal abundance using the code of Nicastro et al.  1999b).
Ionization equilibrium was calculated using CLOUDY (Ferland 1999).
Table 1 gives the equivalent width (EW) of the strongest resonant
lines in four simulations for a cloud at z=0.1 and of equivalent
hydrogen column density of $N_H=10^{20}$ cm$^{-2}$, metal abundance of
$0.3 Z_\odot$, $b=200$ km s$^{-1}$ (the Doppler term $b$ includes also
gas turbulence), and temperatures $T=10^{5.5}, 10^6, 10^{6.5}$ and
$10^7$ K. K$\alpha$ and K$\beta$ resonant transitions from He- and
H-line ions of C, O and Ne produces the strongest features in the
0.2-3 keV band, along with a series of L resonant lines from Mg, S, Si
and Fe.  At $T=10^{5.5}$ K CV and OVII are the dominant ions (their
relative abundance being $>80\%$).  At $T=10^6$K He-like ions of C, O
and Ne, are still dominant, but H-like C is visible. At $T=10^{6.5}$ K
OVIII is the dominant oxygen ion. A relatively strong FeXVII line at
0.825 keV is also visible.  For $T=10^7$ K oxygen is nearly completely
ionized and only a weak L-``forest'' from highly ionized iron is
present in the spectrum around 1 keV (FeXX at 0.967 keV being the
strongest line).  Reducing $b$ by a factor of two will reduce the EW
by $\approx30\%$.  Figure 2 shows how the spectra with $T=10^{6}$ K
and $T=10^{6.5}$ K would be measured by a typical grating (resolution
of $\sim1000$ at 0.3 eV, 500 at 0.5 keV and $\sim300$ at 0.8 keV, and
with similar sensitivity in this band).  We assumed $10^5$ counts
(0.5-2 keV) in each spectrum. With this signal to noise ratio
($\sim10$ per eV) and with a resolution of 1 eV at 0.5 keV OVII and
OVIII lines of EW=0.5-1 eV can be easily detected. In fact, the
minimum EW detectable with a signal to noise of 5 given the above
constraints is of $\sim0.5$ eV (using equation (12) of Perna \& Loeb, 1998).
Ne and Fe lines of similar EW are more difficult to detect, due to the
reduced resolution at increasing energies.  Interestigly, CV and CVI
lines at 0.3-0.4 keV of EW of only 0.3-0.4 eV are also easily
detectable, thanks to the $\sim0.3$ eV resolution of the gratings at
those energies.  Our simulation shows that oxygen of column densities
as low as $N_H=2-4\times10^{16}$ cm$^{-2}$ can be detected in such
high quality spectra at z=0.1. Columns smaller by a factor of
$\approx2$ may be detected at z$\approx 1$, where the OVII and OVIII
lines are redshifted to $\approx0.3$ keV, due again to the improved
resolution of gratings at low energies.

Our calculations adopt a collisional ionisation equilibrium. Whereas,
photoionization from cosmic X-ray and UV backgrounds and from possible
nearby sources like AGN and starburst galaxies, is expected to alter
the relative ionic abundances, making their distributions as a
function of the temperature wider than in the collisional case
(Nicastro et al. 1999a).  The contribution of photoionization to the
gas ionization state is increasingly important for decreasing gas
densities. In this cases, the measurement of the equivalent width at
least three lines of the same element can be used as a temperature
{\it and} density diagnostics.  The contribution of the afterglow
itself to the gas ionization is negligible for distances higher than a
few Mpc. The maximum number of ionizations per event is in fact
$\approx5\times10^{62}$ and assuming conservatively that all the gas
in the IGM has T=$10^6$~K, and that the mean density is
$5\times10^{-7}$ cm$^{-3}$, the radius of the ionized bubble is $\ls3$
Mpc (alse see Perna \& Loeb 1998b).

Finally we note that in principle combination of multiple absorption
systems along a line of sight could complicate the resulting emerging
spectrum, making the identification of the single components more
difficult. OVII and OVIII K$\alpha$ and K$\beta$ lines are by far the
most intense lines expected in a broad temperature range. Several
authors evaluated the distribution of the oxygen line EW per unit
redshift (Hellsten et al. 1998) or the probability to observe a given
line EW per unit redshift (Perna \& Loeb 1998, Jahoda, et al. 2000). 
The results indicate that a random line of sight up to
z=0.3-0.5 should contain the order of one system, or a few systems at
most, with OVII or OVIII of EW$\gs0.5$ eV, easing the line
identification process.

\begin{table*}[ht]
\caption{\bf Equivalent width in eV of the strongest resonant lines}
\begin{tabular}{lcccccc}
\hline
\hline
T (K)      & CV & CVI & OVII & OVIII & NeIX & Fe \\
\hline
$10^{5.5}$ & 0.63-0.36 & --  & 1.12-0.55 &  &  -- & --  \\
$10^{6.0}$ & 0.42-0.16 & 0.23-0.37 & 1.16-0.72 & --  & 0.65-0.17 
& FeIX=0.11\\
$10^{6.5}$ & -- & 0.01-0.02 & 0.43-0.12 & 0.31-0.53 & 0.56-0.14 
& FeXVII=0.31 \\ 
$10^{7.0}$ & -- & -- & -- & 0.02-0.04 & NeX=0.04  
& FeXX=0.10 \\
\hline
\end{tabular}

K$\alpha$ and K$\beta$ lines equivalent width is reported
for the C, O and Ne ions.
\end{table*}

\section {Discussion and Conclusions}

The X-ray band provides a unique opportunity to probe the warm IGM
through the detection of the so called ``X-ray forest''. On the
contrary, UV OVI lines can probe relatively low temperature plasmas
only.  We have shown that afterglows of GRBs can provide the most
effective X-ray beacons at high z, to study absorption features due to
the warm IGM.  A few afterglows per year should be bright enough, two-four
hours after the main event, to produce spectra with 10,000-100,000
counts in the Chandra and XMM gratings, thus allowing the detection of
warm IGM with H equivalent column density of $\gs10^{20}$
cm$^{-2}$. This can then be used to constrain the baryon cosmological
mass density, independent on nucleosynthesis calculations.  Swift,
currently the only planned satellite with a much faster pointing
capability (minutes), will allow the detection of absorption edges and
blended lines from higher density regions ($N_H\gs10^{21}$ cm$^{-2}$)
with the resolution afforded by its CCDs.  Maintaining the Swift
default pointing position (i.e. when the satellite is not observing a
GRB) close to the direction of large structures like the Acquarius
cluster concentration (Markevitch 1999), the Shapley super-cluster,
etc, would increase the probability to observe a GRB behind these high
density structures. This in turn would increase the probability of
detecting the warm IGM component, if it is associated with the
filaments connecting high--density regions as predicted by current
cosmological scenarios (Dav\`e et al 2000).  Detection of
significant absortion in these lines of sight would provide at the
same time a first, direct, {\it a priori} test of the growth of
large--scale structure in the Universe.

More ambitiously, the technology and most of the hardware to put
together a mission with the Swift slew capability, $\sim0.1$ m$^2$
collecting area and high resolution capabilities (such those provided
by gratings or calorimeters), is currently available. The Swift
trigger might be used by such a mission to slew quickly on a GRB
event. Gratings can provide enough resolution to detect oxygen lines
of EW$\gs0.5$ eV at z=0.1-0.3 and, most intriguingly, EW of a few tenth
of eV at z$\sim1$. This will allow to probe oxygen columns as small 
as $10^{16}$ cm$^{-2}$ at such cosmologic redshifts.  Moreover, it
might be worth considering upgrading the slew capabilities of
satellites such as Constellation-X and XEUS. These satellites will
have the throughput to acquire high resolution spectra with the order
of millions of counts from GRB X-ray afterglows, provided that they
can be observed fast enough after the GRB event (i.e. within one
hour).  This may allow one to push the study of the warm IGM to
redshifts $>>1$, through the detection of Fe, Mg, Si and S lines,
opening the way to a {\it direct} test of hierarchical clustering
models for the growth of structures.

\medskip
This research has been partially supported by ASI contract ARS--99--75
and MURST grant Cofin--98--032. We thank A. Comastri, G. Ghisellini,
G.  Matt, C. Norman, P. Padovani, O. Pantano, L. Piro, N. White 
and A. Wolter for useful discussions and an anonymous referee for 
suggestions which improved the presentation.



\begin{references}

Aldcroft, T., Elvis, M., McDowell, J. \& Fiore, F. 1994, ApJ, 437 584

Cavaliere, A., Giacconi, R., \& Menci, N. 2000, ApJL, 528, L77

Cen, R., Ostriker, J.P. 1999, ApJ, 519, L109

Ciardi, B., \& Loeb, A. 2000, ApJ, submitted, astro-ph/0002412

Comastri,A., Setti,G., Zamorani,G., Hasinger,G. 1995, A\&A 296, 1

Costa, E., 1999, A\&AS, 138, 425.


Dav\`e, R. et al. 2000, ApJ, submitted, astro-ph/0007217

Ellison et al., 1999; ApJ, 520, 456

Ferland, G. 1999, CLOUDY version 90.04

Fishman, G.J. \& Meegan, C.A. 1995, A\&ARA, 33, 415

Frontera, F. et al. 2000a, ApJ, in press, astro-ph/9911228

Frontera, F. et al. 2000b, ApJ, in press, astro-ph/0002257

Grossan, 1992, MIT, PhD thesis

Hasinger, G. 1998, Nucl. Phys. B 69/2-3 p. 600

Hellsten, U., Gnedin, N.Y., Miralda-Escud\`e, J. 1998, ApJ, 509, 56

Kirkman, D., \& Tytler, D. 1997, ApJL, 489, L123

Kirkman, D., \& Tytler, D. 1999, ApJL, 512, L5

Jahoda, K., Madejski, G. \& Stahle, C. 2000, \\
http://conxproject.gsfc.nasa.gov/fstjuneinfo.htm

Lamb, D.Q., \& Reichart, D.E. 2000, ApJ, 536, 1

Owens, A., et al. 1998, A\&AL, 339, L37 

Markevitch, M. 1999, ApJL, 522, L13

Nicastro, F., Fiore, F., G.C. Perola, \& Elvis, M. 1999a, ApJ, 512, 184

Nicastro, F., Fiore, F. \& Matt, G. 1999b, ApJ, 517, 108

Perna, R, \& Loeb, A. 1998a, ApJL, 503, L135

Perna, R, \& Loeb, A. 1998b, ApJ, 501, 467

Piro, L. 1999, to appear in ``X-Ray Astronomy '99'', astro-ph/0001436


Savaglio, S., Panagia, N., \& Stiavelli, M. 2000, 
ASP Conf Series, Vol. 3, 2000, astro-ph/9912112

Shapiro, P.R., \& Bahcall, J.N. 1980, ApJ241, 1

Sherman, R.D., \& Silk, J. 1979, ApJL, 234, L9

Songaila \& Cowie 1996, AJ, 112, 335

Tripp, T.M., Savage, B.D., \& Jenkins, E.B. 2000, ApJL, 534, L1

Tripp, T.M., \& Savage, B.D., 2000, ApJ, 542

Wolter, A., Gioia, I.M., Maccacaro, T., Morris, S., \& Stocke, J.T.
1991, ApJ, 369, 314


\end{references}
\end{document}